\documentclass[letterpaper]{ptephy_v1}

\usepackage{amsmath}
\usepackage{graphics}
\usepackage{url}

\usepackage{color}

\def\edth{{\rlap{$\partial$}\raise0.3em\hbox{$-$}}}
\newcommand{\bea}{\begin{eqnarray}}
\newcommand{\eea}{\end{eqnarray}}

\begin{document}

\title{How close can we approach the event horizon of
the Kerr black hole from the detection of  the gravitational quasinormal modes?}


\author{Takashi Nakamura and Hiroyuki Nakano}

\address{Department of Physics, Kyoto University, Kyoto 606-8502, Japan}

\begin{abstract}
Using the WKB method, we show that the peak location ($r_{\rm peak}$)
of the potential, which determines the quasinormal mode frequency
of the Kerr black hole, obeys an accurate empirical relation
as a function of the specific angular momentum $a$
and the gravitational mass $M$. If the quasinormal mode
with $a/M \sim 1$ is observed by gravitational wave detectors,
we can confirm the black-hole space-time around the event horizon,
$r_{\rm peak}=r_+ +O(\sqrt{1-q})$ where $r_+$ is the event horizon radius. 
While if the quasinormal mode is different from that of general relativity,
we are forced to seek the true theory of gravity and/or
face to the existence of the naked singularity.
\end{abstract}

\subjectindex{E31, E02, E01, E38}

\maketitle

\section{Introduction}

Coalescing binary black holes (BHs) form a BH
in numerical relativity simulations~\cite{Pretorius:2005gq, Campanelli:2005dd, Baker:2005vv}.
The BH radiates characteristic gravitational waves (GWs)
with the quasinormal mode (QNM) frequencies which dominates
in the final phase of the merger of two BHs.
QNM will be detected by
the second generation GW detectors such as
Advanced LIGO (aLIGO)~\cite{TheLIGOScientific:2014jea}, 
Advanced Virgo (AdV)~\cite{TheVirgo:2014hva}, 
and KAGRA~\cite{Somiya:2011np,Aso:2013eba}.
QNM is  also one of the targets
for space based GW detectors
such as eLISA~\cite{Seoane:2013qna} and DECIGO~\cite{Seto:2001qf}.
However to confirm the QNM GWs,
we need a sufficiently high signal-to-noise ratio
(see e.g., Refs.~\cite{Berti:2007zu,Nakano:2015uja}).

To calculate GWs from the Kerr BH~\cite{Kerr:1963ud},
we need to solve the Teukolsky equation~\cite{Teukolsky:1973ha}.
The radial Teukolsky equation for gravitational perturbations
in the Kerr space-time which is expressed as 
\begin{eqnarray}
 \Delta^2\frac{d}{dr}\frac{1}{\Delta}\frac{dR}{dr}-VR = -T \,,
\end{eqnarray}
where $T$ is the source and
\begin{eqnarray}
 V = -\frac{K^2}{\Delta}-\frac{2iK\Delta'}{\Delta}+4iK'+\lambda \,,
\end{eqnarray}
with
\begin{eqnarray}
 K = \left(r^2+a^2\right)\,\omega - am \,,
 \label{eq:K}
\end{eqnarray}
where $\Delta = r^2 - 2 M r + a^2$ with $M$ and $a$ being
the mass and the spin parameter, respectively.
In this paper, we use the geometric unit system, where $G=c=1$.
Here, we consider the Kerr metric in the Boyer-Lindquist coordinates as
\bea
 ds^2 &=& 
 - \left( 1 - \frac{2 M r}{\Sigma} \right)  dt^2 
 - \frac{4 M a r ~{\rm{sin}^2 \theta} }{\Sigma} dt d\phi 
 + \frac{\Sigma}{\Delta} dr^2
 \cr &&
 + \Sigma d\theta^2
 + \left( r^2 + a^2 + \frac{2 M a^2 r}{\Sigma} \sin^2 \theta \right)
 \sin^2 \theta d\phi^2 \,,
\eea
where $\Sigma = r^2 + a^2 \cos^2 \theta$.
The constants $m$ and $\lambda$ in the Teukolsky equation
come from  the spin-weighted spheroidal function $Z^{a\omega}_{\ell m}(\theta,\phi)$. 
A prime is the derivative with respect to $r$.
When we consider GWs
emitted by a test particle falling into the Kerr BH,
the source term $T$ diverges as $\propto r^{7/2}$
and the potential $V$ is the long range one, which motivated  
Sasaki and Nakamura~\cite{Sasaki:1981kj,Sasaki:1981sx,Nakamura:1981kk}
to consider the change of the variables and the potential.
Using two functions $\alpha(r)$ and $\beta(r)$ for the moment,
let us define various variables as
\begin{eqnarray}
 X &=& \frac{\sqrt{r^2+a^2}}{\Delta}
 \left(\alpha R+\frac{\beta}{\Delta}R'\right) \,,
 \label{eq:XR} \\
 \gamma &=& \alpha \left(\alpha+\frac{\beta'}{\Delta}\right)
 -\frac{\beta}{\Delta}\left(\alpha'+\frac{\beta}{\Delta^2}V\right) \,,
 \label{eq:gamma} \\
 F &=& \frac{\Delta}{r^2+a^2}\frac{\gamma'}{\gamma} \,, \\
 U_0 &=& V+\frac{\Delta^2}{\beta}
 \left[ \left(2\alpha+\frac{\beta'}{\Delta}\right)'
 -\frac{\gamma'}{\gamma}\left(\alpha+\frac{\beta'}{\Delta}\right)\right] \,,
 \label{eq:U0} \\
 G &=& -\frac{\Delta'}{r^2+a^2}+\frac{r\Delta}{(r^2+a^2)^2} \,,
 \label{eq:G} \\
 U &=& \frac{\Delta U_0}{(r^2+a^2)^2}+G^2+\frac{dG}{dr^*}
 -\frac{\Delta G}{r^2+a^2}\frac{\gamma'}{\gamma} \,.
 \label{eq:U}
\end{eqnarray}
Then, we have a new wave equation for $X$ from the Teukolsky equation as 
\begin{equation}
 \frac{d^2X}{dr^{*2}}-F\frac{dX}{dr^*}-UX=0 \,,
\end{equation}
where $dr^*/dr = (r^2+a^2)/\Delta$.

We define $\alpha$ and $\beta$ by
\begin{eqnarray}
 \alpha &=& A-\frac{iK}{\Delta} B \,, \\
 \beta &=& \Delta B \,,
\end{eqnarray}
where
\begin{eqnarray}
 A &=& 3iK'+\lambda+\Delta P \,, \\
 B &=& -2iK+\Delta'+\Delta Q \,,
\end{eqnarray}
with
\begin{eqnarray}
 P &=& \frac{r^2+a^2}{gh}\left(\left(\frac{g}{r^2+a^2}\right)'h\right)' \,, \cr
 Q &=& \frac{(r^2+a^2)^2}{g^2h}\left(\frac{g^2h}{(r^2+a^2)^2}\right)' \,.
\end{eqnarray}
Here, $g$ and $h$ are free functions under the restrictions
to guarantee the convergent source term and the short range potential
which is given by
\begin{eqnarray}
g = {\rm const} \,, \quad h={\rm const} \,,
\end{eqnarray}
for $r^* \to -\infty$, and
\begin{eqnarray}
g = {\rm const} +O(r^{-2}) \,, \quad h={\rm const}+ O(r^{-2}) \,,
\label{eq:gh_inf}
\end{eqnarray}
for $r^* \to +\infty$.
In Refs.~\cite{Sasaki:1981sx, Sasaki:1981kj},
$h$ and $g$ are adopted as
\begin{equation}
h=1 \,, \quad g=\frac{r^2+a^2}{r^2} \,.
\end{equation}
Defining a new variable $Y$ by $X=\sqrt{\gamma}~Y$, we have
\begin{eqnarray}
 \frac{d^2Y}{dr^{*2}}+\left( \omega^2-V_{\rm SN} \right) Y =0 \,,
\end{eqnarray}
where
\begin{eqnarray}
 V_{\rm SN}=\omega^2+U
 -\left[ \frac{1}{2}\frac{d}{dr^*}
 \left(\frac{1}{\gamma}\frac{d\gamma}{dr^*}\right)
 -\frac{1}{4\gamma^2}\left(\frac{d\gamma}{dr^*}\right)^2\right] \,.
\end{eqnarray}

In our previous paper~\cite{Nakamura:2016gri},
it was shown that this $V_{\rm SN}$ has
double peaks for $q = a/M > 0.8$
to refuse the approach to determine the complex frequency
of QNMs from the peak location $r_{\rm peak}$
which is real-valued, of the absolute value of $V_{\rm SN}$.
Our new $g$ is defined by
\begin{equation}
g=\frac{r(r-a)}{(r+a)^2} \,.
\label{eq:new_g}
\end{equation}
The above new $g$ seems to violate the needed dependence
for $r \to \infty$. However Eq.~\eqref{eq:gh_inf} tells
only the sufficient condition but not the necessary one.
In reality, our new $V_{\rm NNT}$~\footnote{To distinguish
the original $V_{\rm SN}$ from the new SN equation,
we use $V_{\rm NNT}$ from now on.}
is confirmed to be short-ranged, that is, $V_{\rm NNT}$ is
in promotion to $1/r^2$ for $r \to \infty$ and becomes the
Regge-Wheeler potential~\cite{Regge:1957td} for $a=0$ as $V_{\rm SN}$.

After adopting the new potential $V_{\rm NNT}$,
what we will do is to ask which part of the Kerr metric
determines the QNMs for given $0.8 < q < 1$.
Conversely, if the QNM GWs are observed,
which part of the Kerr BH we can say being confirmed,
which is the main theme of this paper. 

The essential technique  to determine the QNMs by using the WKB method was proposed
by Schutz and Will~\cite{Schutz:1985zz}
 for the Schwarzschild BH using 
the Regge-Wheeler potential $V_{\rm RW}$~\cite{Regge:1957td,Zerilli:wd}. 
They approximated $V_{\rm RW}$
near its peak radius at $r_0 \approx 3.28M$ as 
\begin{eqnarray}
 V_{\rm RW}(r^*) = V_{\rm RW}(r^*_0)
 +\frac{1}{2} \left. \frac{d^2V_{\rm RW}}{dr^{*2}}
 \right|_{r^*=r^*_0}(r^*-r^*_0)^2 \,,
\end{eqnarray}
where $r^*_0=r_0+2M\ln(r_0/2M-1)$.
The QNM frequencies are expressed as
\begin{equation}
 (\omega_r+i\omega_i)^2
 = V_{\rm RW}(r^*_0)-i\left(n+\frac{1}{2}\right)
 \sqrt{-2 \left. \frac{d^2V_{\rm RW}}{dr^{*2}} \right|_{r^*=r^*_0}} \,.
\label{eq:WKB_RW}
\end{equation}
with $n=0,\,1,\,2, \cdot\cdot\cdot$.
As for the accuracy of fundamental $n=0$ QNM frequency with $\ell=2$, 
the errors of the real ($\omega_r = {\rm Re}(\omega)$)
and the imaginary ($\omega_i = {\rm Im}(\omega)$) parts
are $7\%$ and $0.7\%$, respectively,
compared with the numerical results of
Chandrasekhar and Detweiler~\cite{Chandrasekhar:1975zza}.
This suggests that for the fundamental QNM of the $a=0$ case,
the space-time of a Schwarzschild BH around $r \approx 3.28M$
is confirmed through the detection of the QNM GWs.
The word ``around'' has two meaning that the GW
cannot be localized due to the equivalence principle
and the imaginary part of the QNMs is determined
by the curvature of the potential,
which reflects the space-time structure of the Schwarzschild BH
around $r \approx 3.28M$.

In our previous papers~\cite{Nakamura:2016gri, Nakano:2016a},
we succeeded in doing similar procedures up to $q=0.98$
for the Kerr BH with the Detweiler potential~\cite{Detweiler:1977gy}
(see also Ref.~\cite{Chandrasekhar:1976zz}).
However, above $q=0.98$,
we could not derive the consistent QNMs
in the method to use $r_{\rm peak}$ of the absolute value of the potential.

\begin{figure}[!ht]
\begin{center}
 \includegraphics[width=0.5\textwidth,clip=true]{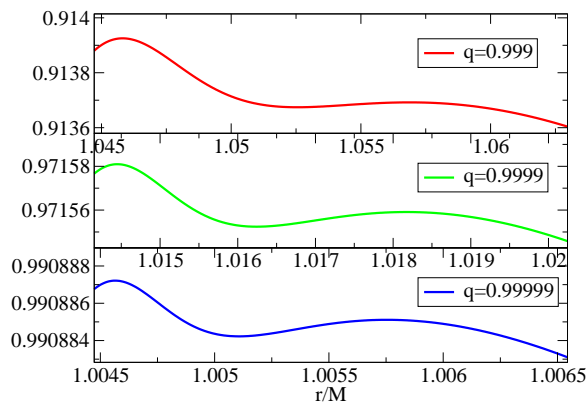}
\end{center}
 \caption{$|V_{\rm NNT}|$ with ($\ell=2,\,m=2$)
 for $q=0.999$ (top), $0.9999$ (middle) and $0.99999$ (bottom)
 with respect to $r/M$. We will calculate the QNM frequencies
 by using the strong peak for each $q$.}
 \label{fig:NNT_ptl}
\end{figure}

In Fig.~\ref{fig:NNT_ptl}, we show that new Sasaki-Nakamura potential $V_{\rm NNT}$
has only a single strong peak to allow us to identify $r_{\rm peak}$
up to $q \rightarrow 1$ successfully.
Here, the QNM frequencies have been calculated
accurately by the Leaver's method~\cite{Leaver:1985ax}.
It is noted that when we consider an additional function $a (M-a)/r^2$
for Eq.~\eqref{eq:new_g}, the peak location $r_{\rm peak}$ changes
only $0.4\%$ and $0.06\%$ for $q=0.98$ and $0.9999$, respectively,
which show that the results do not depend on the choice
of $g$ significantly.

\section{Result}

\begin{figure}[!ht]
\begin{center}
 \includegraphics[width=0.5\textwidth,clip=true]{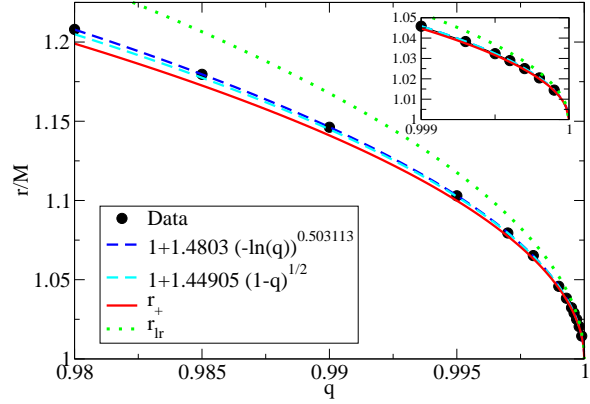}
\end{center}
 \caption{Black dots denote the locations of the maximum of
 the absolute value of $|V_{\rm NNT}|$ with ($\ell=2,\,m=2$)
 for various spin parameters $q=a/M$.
 The blue and light blue dashed curves are the fitting curve
 and the curve derived analytically. 
 The red curve is the event horizon $r_{+}$,
 and the inner light ring radius $r_{\rm lr}$
 is presented by the green dotted curve.}
 \label{fig:location_new}
\end{figure}

First, we calculate the peak location of $V_{\rm NNT}$
for various spin parameters $q$,
and show various curves in Fig.~\ref{fig:location_new}.
Black dots are obtained by finding the maximum of the absolute value
of the new Sasaki-Nakamura potential $|V_{\rm NNT}|$ with ($\ell=2,\,m=2$),
where the contribution from the imaginary part is small.
In practice, $|{\rm Im} (V_{\rm NNT})/{\rm Re} (V_{\rm NNT})|$
is $9\%$ ($q=0.98$), $7\%$ ($q=0.99$), $2\%$ ($q=0.999$),
$0.7\%$ ($q=0.9999$) and $0.2\%$ ($q=0.99999$).

The blue curve in Fig.~\ref{fig:location_new} is the fitting curve with
\bea
r_{\rm fit}/M = 1+1.4803 \,(-\ln q)^{0.503113} \,,
\eea
which is derived from 13 data between $q=0.98$ and $q=0.9999$
yielding the correlation coefficient of $0.999995$
and the chance probability of  $7.4\times 10^{-29}$.
The red and green dotted curves present the event horizon radius,
\bea
 r_{+}/M &=& 1 + \sqrt{1-q^2} 
 \cr 
 &=& 1 + \sqrt{2}\,(1-q)^{1/2}
 + O((1-q)^{3/2}) \,,
\eea
and the inner light ring radius~\cite{Bardeen:1972fi},
\bea
 r_{\rm lr}/M &=& 2 + 2\cos\left[\frac{2}{3}
 \cos^{-1}\left(-q\right)\right] 
 \cr
 &=& 1 + \frac{2\sqrt{2}}{\sqrt{3}}\,(1-q)^{1/2}
 + O(1-q) \,.
\eea
The latter radius is evaluated in the equatorial ($\theta=\pi/2$) plane.
It is noted that there are various studies
on the relation between the QNMs
and the orbital frequency of the light ring orbit
(see a useful lecture note~\cite{Berti:2014bla}).  

The light blue dashed curve in Fig.~\ref{fig:location_new}
is obtained as follows.
Here, we introduce a fitting curve to evaluate $\lambda$
by using ${}_sA_{\ell m}$ which is a constant defined in Eq.~(25)
of Ref.~\cite{Berti:2009kk}, as
\bea
 {}_{-2}A_{22} = 0.545652 + (6.02497+1.38591 \,i) (-\ln q)^{1/2} \,,
\eea 
where $\lambda$ is related to ${}_sA_{\ell m}$ as
$\lambda = {}_sA_{\ell m} - 2am\omega + a^2\omega^2$.
And also, the ($n=0$) QNM frequency with ($\ell=2,\,m=2$)
is described by Ref.~\cite{Hod:2008zz} as
\bea
M \omega = \frac{M q}{r_+}-\frac{i}{4} \frac{r_+ - M}{r_+}
\,.
\eea
Then, using $q=1-\epsilon$ and expanding $V_{\rm NNT}$
with respect to $\epsilon$, we estimate the $r_{\rm peak}$
analytically. Instead of finding the peak location of $|V_{\rm NNT}|$,
we derive the location of $dV_{\rm NNT}/dr^*=0$.
The location has a $O(\sqrt{\epsilon})$ term
which is consistent with $r_{\rm fit}$ because 
$(-\ln(1-\epsilon))^{1/2}=\sqrt{\epsilon}+O(\epsilon^{3/2})$, and
the result is
\bea
r_{\rm ana}/M = 1 + 1.44905 \,(1-q)^{1/2} \,.
\eea
Here, we have ignored the tiny imaginary contribution of $-0.020157\,i$.
It should be noted that a different choice of $g$ makes
a difference of $O((1-q)^{1/2})$ in the estimation of the peak location.
We will discuss the detail in our future work.

\begin{figure}[!ht]
\begin{center}
 \includegraphics[width=0.5\textwidth,clip=true]{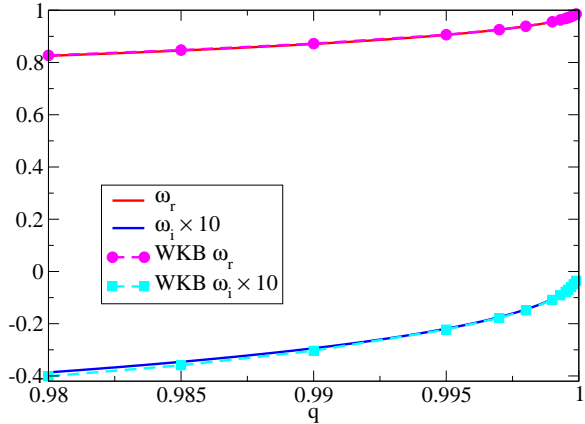}
\end{center}
 \caption{The real and imaginary parts of the fundamental ($n=0$)
 QNM frequencies with $V_{\rm NNT}(\ell=2,\,m=2)$
 evaluated for various spin parameters $q=a/M$.
 The exact frequencies $\omega_r$ and $\omega_i$
 are from Refs.~\cite{Berti:2005ys,BertiQNM},
 and the Leaver's method~\cite{Leaver:1985ax}.}
 \label{fig:frequency_new}
\end{figure}

\begin{figure}[!ht]
\begin{center}
 \includegraphics[width=0.5\textwidth,clip=true]{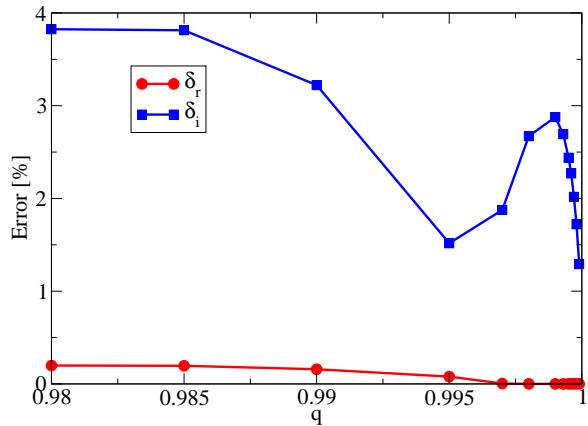}
\end{center}
 \caption{Absolute value of relative errors for the real and imaginary part
 of the QNM frequencies with $V_{\rm NNT}(\ell=2,\,m=2)$,
 $\delta_{\rm r}=|({\rm WKB}~\omega_r)/\omega_r-1|$
 and $\delta_{\rm i}=|({\rm WKB}~\omega_i)/\omega_i-1|$
 between the exact value and that of the WKB approximation
 in Fig.~\ref{fig:frequency_new}.}
 \label{fig:error_new}
\end{figure}

To confirm our analysis, we present the QNM frequencies
via the WKB approximation in Fig.~\ref{fig:frequency_new}.
The errors in the real and imaginary parts of the QNM frequencies
are plotted in Fig.~\ref{fig:error_new}.

\section{Discussion}

There is the ergoregion in the Kerr BH
where the timelike Killing vector turns out
to be spacelike. The boundary of the ergoregion
is given by
\bea
 r_{\rm ergo}(\theta)/M = 1 + \sqrt{1-q^2\cos^2 \theta} \,,
\eea
which is called the ergosphere.
Note here that $r_{\rm ergo}(0)=r_+$ and $r_{\rm ergo}(\pi/2)=2M$.
The physical origin of the Penrose mechanism~\cite{Penrose:1969pc}
and the Blanford-Znajek mechanism~\cite{Blandford:1977ds} is the eroregion,
which enables the extraction of the rotational energy of the Kerr BH.
There are many papers using these mechanisms
while the confirmation of the existence
of the ergoregion has not been done observationally.
The detection of QNM GWs, for example,
for $q=0.9999$ can confirm its existence.
Let us define the covering solid angle $4\pi C$ of the ergoregion
for the given sphere of the radius $r_{\rm peak}$.
Then, $C$ is given by
$C=\cos \theta_{\rm m}$ with  $r_{\rm peak}/M=1+\sqrt{1-q^2\cos^2\theta_{\rm m}}$. 
Then if, for example, the QNM with $q=0.9999$ is observed by GW detectors,
we can confirm the space-time around $r=1.01445M$
covering  $99.9996\%$ of the ergoregion.
Here, the horizon radius for $q=0.9999$ is $1.01414M$.
Therefore, the space-time at only $1.0003$ times the event horizon can be confirmed.
If the QNM is confirmed to be different from that of general relativity
by the detection of the GWs, a very serious problem is raised
since the Kerr BH is the unique solution
of the stationary vacuum solution of the Einstein equation
under the assumption of the cosmic
censorship~\cite{Israel:1967wq,Carter:1971zc,Robinson:1975bv}.
The Einstein equation and/or the cosmic censorship are wrong.
In the former case, the true theory of gravity should be determined
to be compatible with the data of QNMs.
In the latter case, we face the existence of the naked singularity,
which requires a new physics law
possibly related to quantum gravity.

As for the detection rate,
the population  Monte Carlo  simulation
by Kinugawa et al.~\cite{Kinugawa:2014zha,Kinugawa:2015nla,Kinugawa:2016mfs}
showed that for the Pop III binary BHs, $0.43\%$ have the final $q > 0.98$ 
in their standard model,
in which they adopted the various parameters and functions for 
the Pop I stars like the sun  except for the initial mass function.
Since the Pop III star which is the first star in our universe without metal
that has  atomic number larger than carbon has not been observed
so that these parameters are highly unknown.
This suggests that the percentage of BHs with $q > 0.98$ 
can be either larger or smaller than $0.43\%$ .
While the massive BHs from Pop I and Pop II BH binaries
are also expected~\cite{Dominik:2014yma}
although the spin parameter data are not available from their paper at present.
This suggests that  the second generation detectors might detect
the QNM GWs of such BH with $q> 0.98$.
The third generation detectors such as the Einstein Telescope
(ET)~\cite{Punturo:2010zz} will increase the detection
number $\sim 1000$ larger.  Another possibility is QNM GWs from very massive BHs of mass
$\sim 10^4M_\odot$ and $\sim 10^7M_\odot$ for DECIGO~\cite{Seto:2001qf} 
and eLISA~\cite{Seoane:2013qna}, respectively. 

In conclusion, the present and the future GW detectors
with the frequency $\sim 10^{-3}\,$Hz to $100\,$Hz,
would observe the very strong gravity almost at the event horizon 
radius to clarify the true theory of gravity.

\section*{Acknowledgment}

~~~We thank Takahiro Tanaka for very useful comments.
This work was supported by MEXT Grant-in-Aid for Scientific Research
on Innovative Areas,
``New Developments in Astrophysics Through Multi-Messenger Observations
of Gravitational Wave Sources'', No.~24103006 (TN, HN) and
by the Grant-in-Aid from the Ministry of Education, Culture, Sports,
Science and Technology (MEXT) of Japan No.~15H02087 (TN).


\end{document}